\documentclass[12pt]{article}
\usepackage[colorlinks=true                                                                  ,urlcolor=blue                                                                               ,anchorcolor=blue                                                                            ,citecolor=blue                                                                              ,filecolor=blue                                                                              ,linkcolor=blue                                                                              ,menucolor=blue                                                                              ,pagecolor=blue                                                                              ,linktocpage=true                                                                            ,pdfproducer=medialab                                                                       ,pdfa=true                                                                                 ]{hyperref} 
\usepackage{xspace}
\usepackage{graphicx}
\usepackage{amsmath}
\usepackage{xcolor}
\usepackage{float}
\usepackage{comment}
\usepackage{subcaption}
\usepackage{textcomp}
\usepackage{placeins}
\usepackage{bm}
\usepackage{pstricks}
\usepackage{color}
\usepackage{braket}
\usepackage{slashed}
\usepackage{soul}
\usepackage{geometry}
\usepackage{cite}

\usepackage{titlesec}
\titlespacing*{\section}{0pt}{0.9\baselineskip}{0.5\baselineskip}

\newcommand{\rescaletwoplots}{0.49\textwidth}

\newcommand{\geneva}{\textsc{Geneva}\xspace}
\newcommand{\pythiaEight}{\textsc{Pythia8}\xspace}
\newcommand{\dire}{\textsc{Dire}\xspace}
\newcommand{\sherpa}{\textsc{Sherpa}\xspace}

\newcommand{\df}{\mathrm{d}}
\newcommand{\sss}{\mathchoice %                                                                                                                                                            
  {\displaystyle} %                                                                                                                                                                        
  {\displaystyle} %                                                                                                                                                                        
  {\scriptscriptstyle} %                                                                                                                                                                   
  {\scriptscriptstyle} %                                                                                                                                                                   
}
\newcommand{\MAX}{{\sss\rm max}} % maximum                                                                                                                                                 
\newcommand{\MIN}{{\sss\rm min}} % minimum                                                                                                                                                 

\newcommand{\DS}{\displaystyle} % fix formulae in arrays and fractions                                                                                                                     
\renewcommand{\L}{\left}
\newcommand{\R}{\right}

\newcommand\pubdate{\today}

\def\Title#1{\begin{center} {\Large #1 } \end{center}}
\def\Author#1{\begin{center}{ \sc #1} \end{center}}
\def\Address#1{\begin{center}{ \it #1} \end{center}}

\newcommand\pubblock{\rightline{\begin{tabular}{l}  \\ % Author's note number [if you need to add one] goes here
         \pubdate  \end{tabular}}}
\newenvironment{Abstract}{\begin{quotation}  }{\end{quotation}}
\newenvironment{Presented}{\begin{quotation} \begin{center} 
             PRESENTED AT\end{center}\bigskip 
      \begin{center}\begin{large}}{\end{large}\end{center} \end{quotation}}
\begin{document}
\begin{titlepage}
 \pubblock
\vfill
\Title{\geneva Monte Carlo: status and new developments}
\vfill
\Author{Giulia Marinelli}
\Address{Università degli Studi di Milano-Bicocca \& INFN, \\
Piazza della Scienza 3, Milano 20126, Italy}
\vfill
\begin{Abstract}
We review the \geneva Monte Carlo framework, that combines three theoretical tools used for QCD precise predictions into a single structure.
In this talk we highlight its main features, discussing some new improvements involving both colour singlet productions, as well as for the production of final states with heavy coloured partons and jets.
\end{Abstract}
\vfill
\begin{Presented}
DIS2023: XXX International Workshop on Deep-Inelastic Scattering and
Related Subjects, \\
Michigan State University, USA, 27-31 March 2023 \\
     \includegraphics[width=9cm]{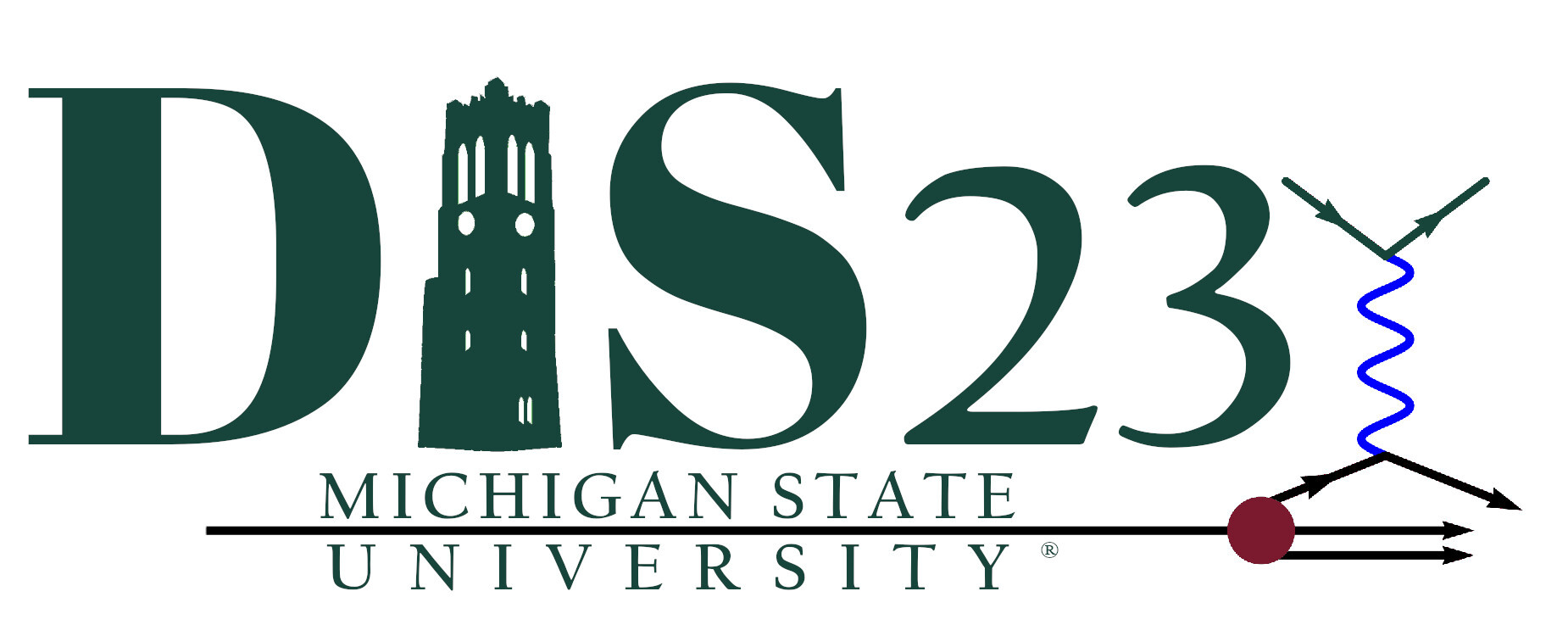}
\end{Presented}
\vfill
\end{titlepage}

\section{The GENEVA framework}
\geneva, which stands for GENerate EVents Analytically, is a Monte Carlo event generator which combines the fixed-order (FO) predictions with the resummed ones, and then merges them with a parton shower (PS).
The aim of \geneva is to obtain fully exclusive hadronized events by combining these three descriptions, while maintaining the perturbative accuracy of a higher-order calculation.
It consistently improves the perturbative accuracy away from the fixed-order regions and provides a systematic estimate of theoretical perturbative uncertainties on an event-by-event basis.
It provides fully differential fixed-order calculations up to next-to-next-to-leading order (NNLO), which are then combined with higher-order resummation in the $N$-jettiness resolution variable.
Currently, for colour singlet processes, the resummation is carried out up to next-to-next-to-next-to-leading logarithmic (N$^3$LL) accuracy for zero-jettiness ($\mathcal{T}_0$) and to next-to-leading logarithmic (NLL) accuracy for one-jettiness ($\mathcal{T}_1$), 
through Soft Collinear Effective Theory (SCET) or transverse momentum resummation.
The resulting parton-level events are further combined with parton shower,
hadronization and multiple particle interaction (MPI) simulations. 
\geneva employs the infrared-safe resolution variables $\mathcal{T}_0$ and $\mathcal{T}_1$ to classify events as having 0- ($\Phi_0$), 1- ($\Phi_1$) or 2- ($\Phi_2$) jets. 
This classification is done by comparing the value of the resolution variables with two cutoffs for each configuration generated. 
Emissions below the resolution cutoff, $\mathcal{T}_N < \mathcal{T}^{\rm cut}_N$, are considered unresolved and are integrated over.
All the configurations with $\mathcal{T}_0 < \mathcal{T}_0^{\rm cut}$ are $\Phi_0$ events,
the ones with $\mathcal{T}_0 > \mathcal{T}_0^{\rm cut}$ but $\mathcal{T}_1 < \mathcal{T}_1^{\rm cut}$ are $\Phi_1$ events and the remaining configurations with $\mathcal{T}_1 > \mathcal{T}_1^{\rm cut}$ are $\Phi_2$ events.

A fundamental role in the \geneva Monte Carlo event generator is played by the splitting functions, which are needed for the mappings used to project the configurations into the $N$-jet bins and to ensure that all the terms in the Monte Carlo cross section are fully differential in $\Phi_N$.
One of the recent improvements is the on-the-fly evaluation of the splitting functions.
Consider a generic splitting $N \to N+1$, the relative splitting function $\mathcal{P}(\Phi_{N+1})$ is defined as 
defined as 
\begin{equation}
  \label{eq:splittingprob}
    \small{
   \mathcal{P} \L( \Phi_{N+1} \R) = \frac{\DS f_{kj} \L( \Phi_N, \mathcal{T}_N, z \R)}{\DS \sum_{k'=1}^{N+2}
    \int_{z_\MIN^{k'} \L(\Phi_N,\mathcal{T}_N \R) }^{z_\MAX^{k'}  \L( \Phi_N, \mathcal{T}_N \R) } \df z'
    J_{k'}  \L( \Phi_N, \mathcal{T}_N, z' \R) I_\phi^{k'} \L( \Phi_N, \mathcal{T}_N, z' \R) \sum_{j'=1}^{n_{\rm split}^{k'}} f_{k'j'}  \L( \Phi_N,\mathcal{T}_N, z' \R)}}\,,   
\end{equation}
where $f_{kj}$ are generic functions based on the Altarelli Parisi splitting functions (depending whether the radiation is in the initial or final state), $z$ is an energy ratio and $\phi$ an azimuthal angle.
These variables are needed besides $\mathcal{T}_N$ to define the a splitting $\Phi_N \to \Phi_{N+1}$.
In Eq.~\eqref{eq:splittingprob}, $J$ is the Jacobian of the change of variables and the $I_{\phi}$ is the integral in $\phi$.
Under the assumption that the Jacobian does not depend on the $\phi$ variable (which holds true for the splittings used in the \geneva framework), the integral in the denominator is computed for each configuration generated, considering the 
appropriate limits of integrations in $z$ and $\phi$.
More details and a comprehensive study on this development can be found in Refs.~\cite{Gavardi:2023oho,Alioli:2023har}.
Another new development in the \geneva framework is the extension of the shower interface, designed for \pythiaEight~\cite{Sjostrand:2004ef}, to include other parton showers such as \sherpa~\cite{Gleisberg:2008ta,Sherpa:2019gpd,Schumann:2007mg} and \dire~\cite{Hoche:2015sya}.
These parton showers differ mainly in their choice of the evolution variable, which, along with the starting scale of the shower matched to the FO and resummed result, determines the extent to which the parton shower explores the phase space beyond the strict soft and collinear limits.
Investigating the effects of different parton showers allows to estimate in a realistic way the uncertainty of the shower matching.

\section{Status of the GENEVA event generator }
Until now, numerous colour singlet processes have been implemented in the \geneva framework.
One of the earliest process has been the Drell-Yan, first with the $N$-jettiness
resummation~\cite{Alioli:2015toa}, and later with the $q_T$ one~\cite{Alioli:2021qbf}.
We then implemented the Higgsstrahlung process~\cite{Alioli:2019qzz} and the NNLL$'$ resummed two-jettiness distribution for decays of the Higgs boson to $b \bar{b}$ and $gg$~\cite{Alioli:2020fzf}.
Also the photon pair and $W \gamma$ production are available~\cite{Alioli:2020qrd,Cridge:2021hfr}.
Recently, we presented the $Z$ boson pair production~\cite{Alioli:2021egp} and the zero-jettiness resummation for $t \bar{t}$ production~\cite{Alioli:2021ggd}.

Some of the latest processes implemented in \geneva involve colour singlet production with gluons in the initial state.
Specifically, we studied single and double Higgs boson production, see Refs.~\cite{Alioli:2023har,Alioli:2022dkj}.
Both processes are considered in the infinite top-quark mass limit ($m_t \to \infty$), where the top-quark is treated as infinitely heavy and is integrated out.
Apart from the current interest in these processes within the particle physics community, their implementation allowed us to directly test the new developments discussed above.

\begin{figure}[t]                                                                                                                                                                       
  \begin{center}
    \includegraphics[width=\rescaletwoplots]{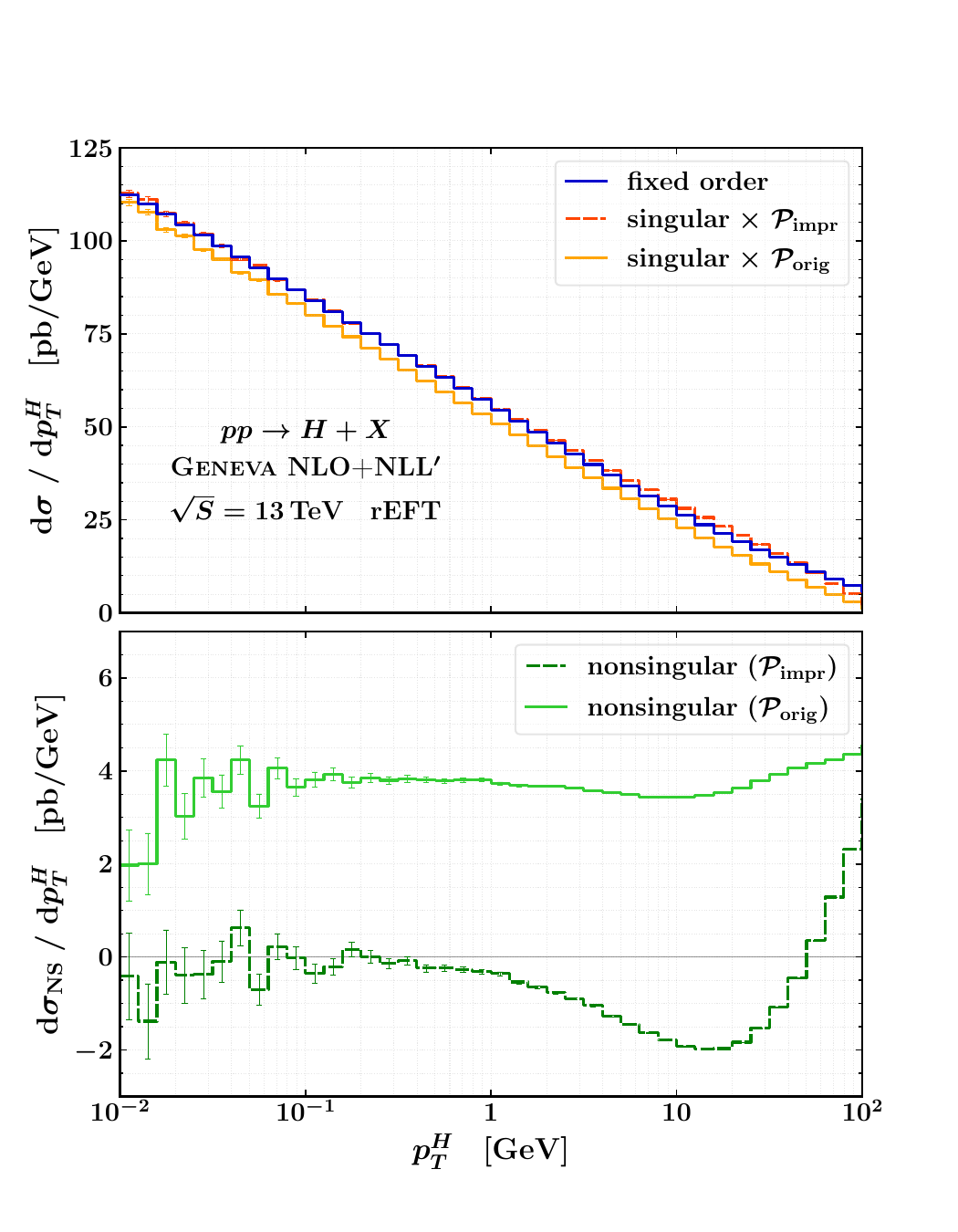}
    \includegraphics[width=\rescaletwoplots]{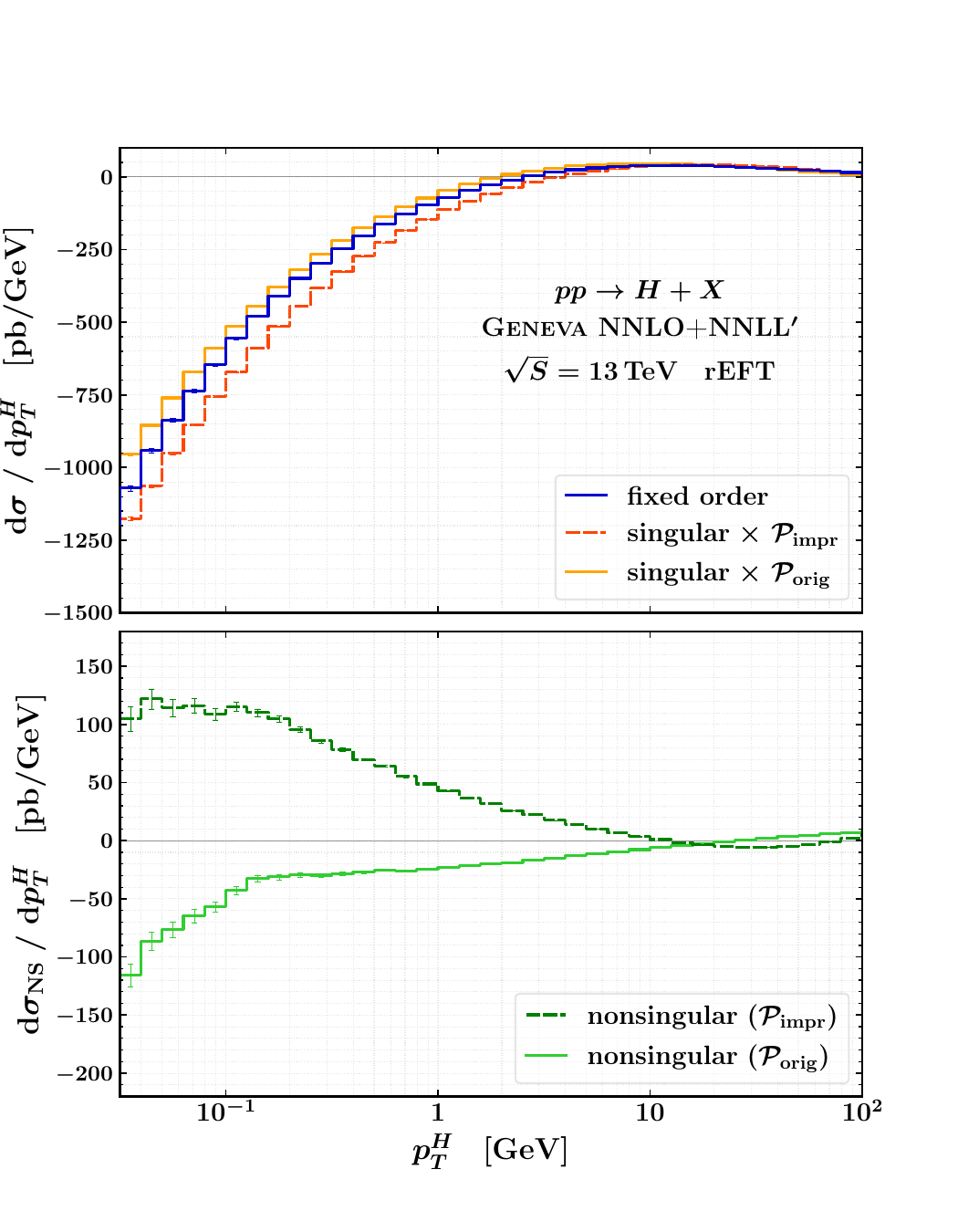}
  \end{center}
  \caption{\label{fig:singlehiggsSplitting}Comparison of the fixed-order, singular, and nonsingular distributions at NLO+NLL$'$ (left) and NNLO+NNLL$'$ (right) for $p_T^H$, both for the original and improved versions of the splitting function implementation.}
\end{figure}

To present the effects of the improved splitting function implementation, in Fig.~\ref{fig:singlehiggsSplitting} we show the comparison between the various contributions of the \geneva cross section at NLO+NLL$'$ (left panel) and NNLO+NNLL$'$ (right panel) for the transverse momentum of the Higgs boson, both for the original and improved version of the splitting function implementation.
The soft and collinear limit for the $\mathcal{P}_{0 \to 1}$ (left panel) is now correctly reproduced as can be
seen by the nonsingular distribution converging to zero.
The same limit for the $\mathcal{P}_{1 \to 2}$ (right panel) is improved, but it appears to miss a single logarithmic contribution. Indeed, the nonsingular distribution converges to a nonzero constant at low value of the transverse momentum.

\begin{figure}[t]                                                                                                                                                                       
  \begin{center}
    \includegraphics[width=\rescaletwoplots]{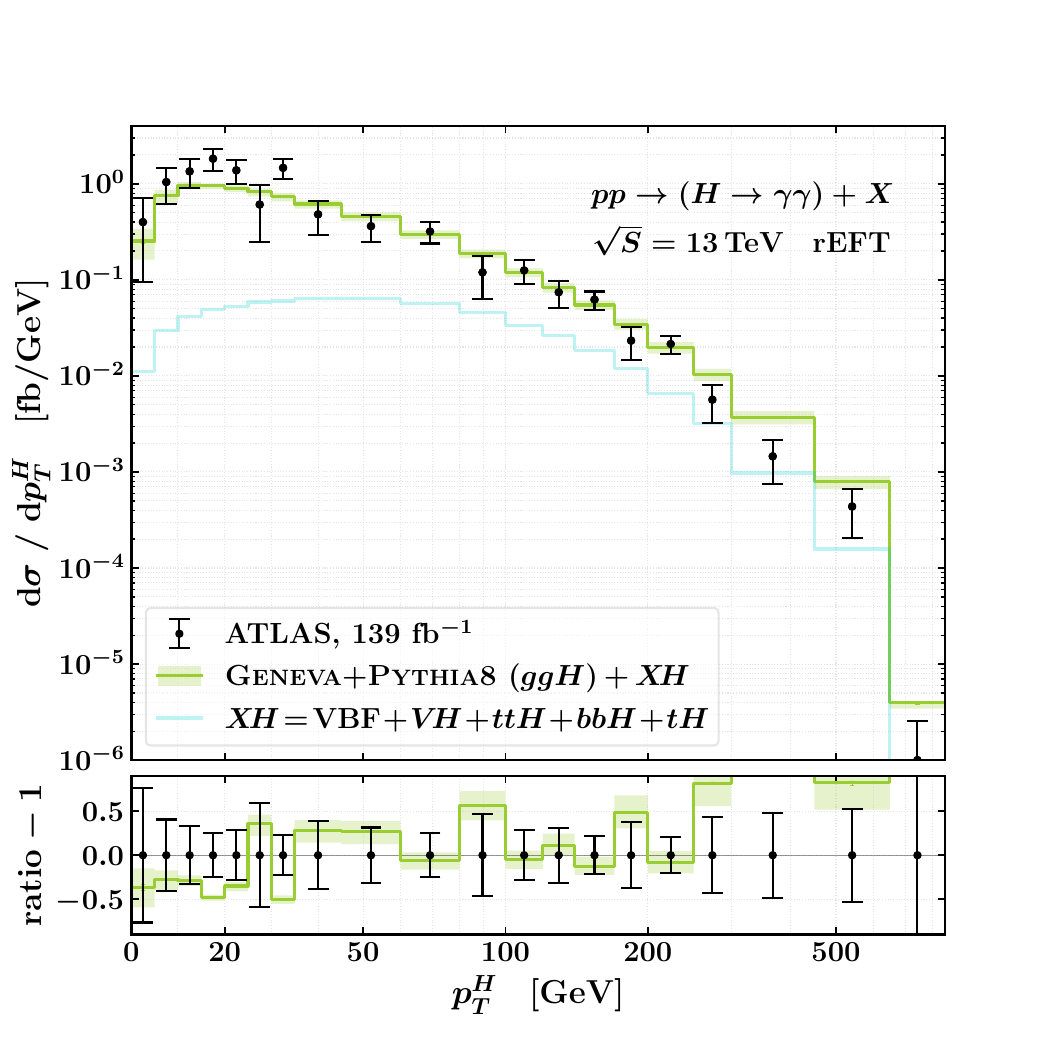}
    \includegraphics[width=\rescaletwoplots]{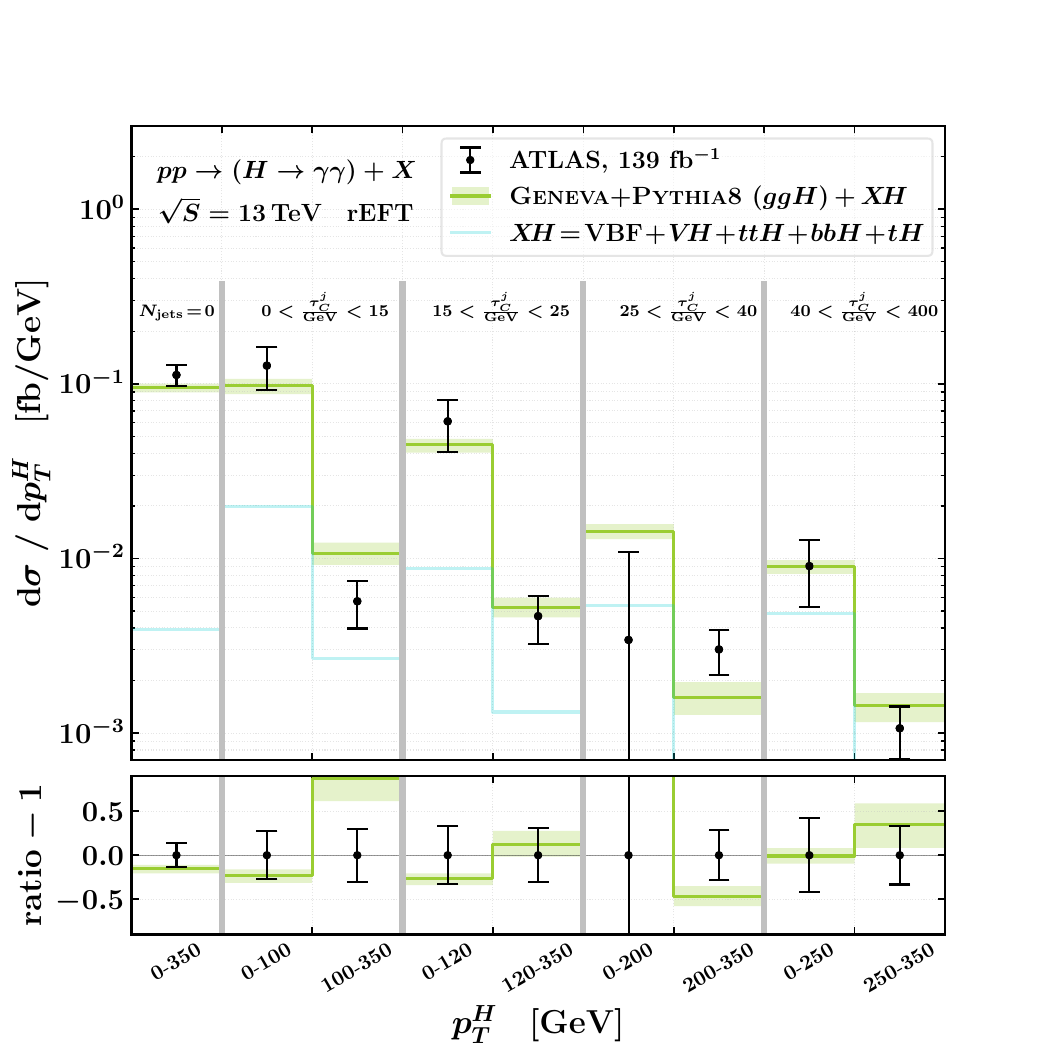}
  \end{center}
  \caption{\label{fig:singlehiggs}Comparison of the ATLAS data~\cite{ATLAS:2022fnp} with the \geneva + \pythiaEight results for the transverse momentum of the Higgs boson (left), also in bins of $\tau^{j_1}_C$ (right).}
\end{figure}

In Fig.~\ref{fig:singlehiggs}, we show \geneva + \pythiaEight (QCD+QED shower, including MPI) result for the transverse momentum of the Higgs boson, compared with the latest experimental results for the Higgs boson inclusive and differential cross sections in the $H \to \gamma \gamma$ decay channel.
We use matrix elements computed in the infinite top-quark mass limit and rescaled in the rEFT scheme.
The contributions from other Higgs boson production modes (named as XH) are included by summing them to the \geneva results for the gluon-fusion channel alone. The XH distributions are obtained from the plots in ATLAS and CMS publications.
We find overall good agreement between the \geneva predictions and the measurements. There is a slight deviation in the peak region and a more pronounced discrepancy in the tail of the distribution,
where the infinity top-quark mass limit approximation is less accurate.
Notably, the \geneva results consider
the 7-points scale variations.

\begin{figure}[H]                                                                                                                                                                       
  \begin{center}
    \includegraphics[width=\rescaletwoplots]{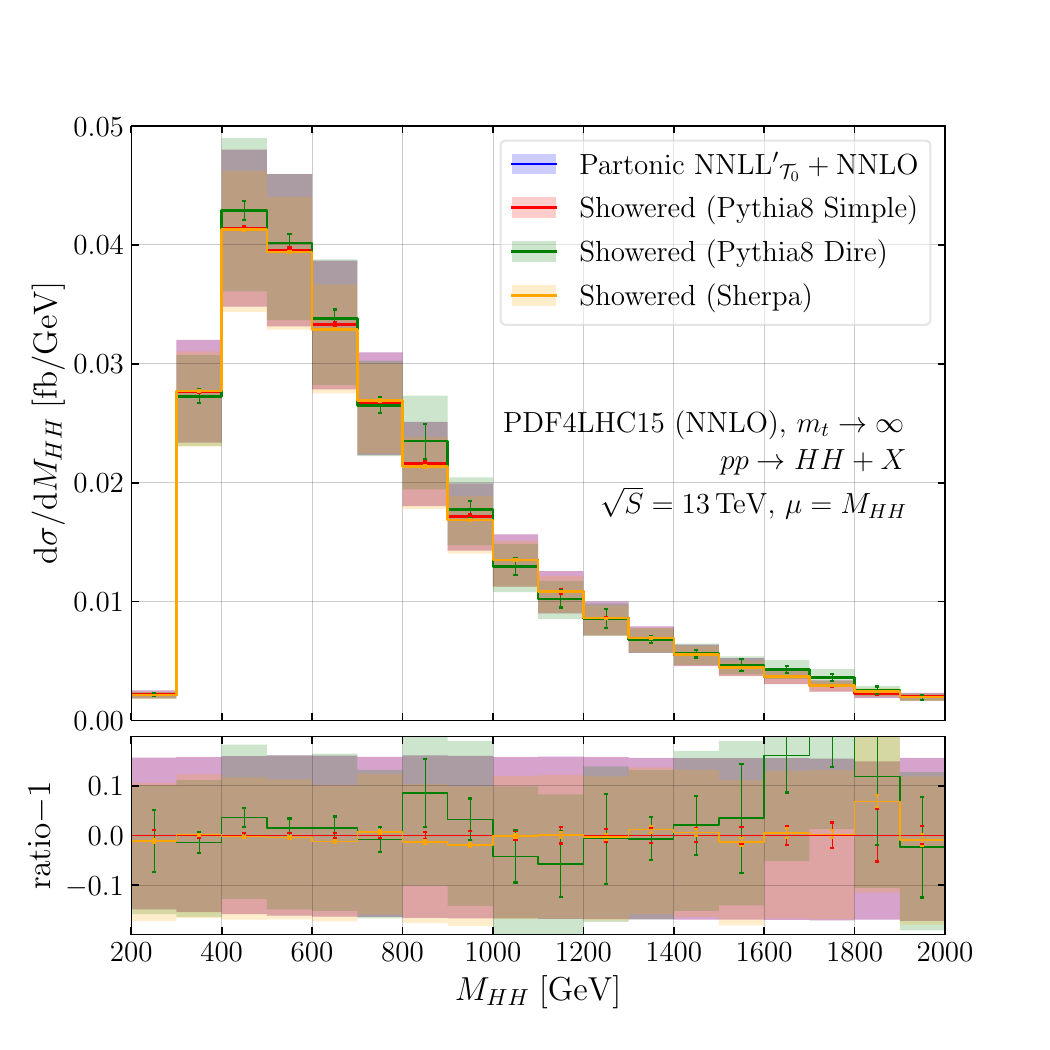}
    \includegraphics[width=\rescaletwoplots]{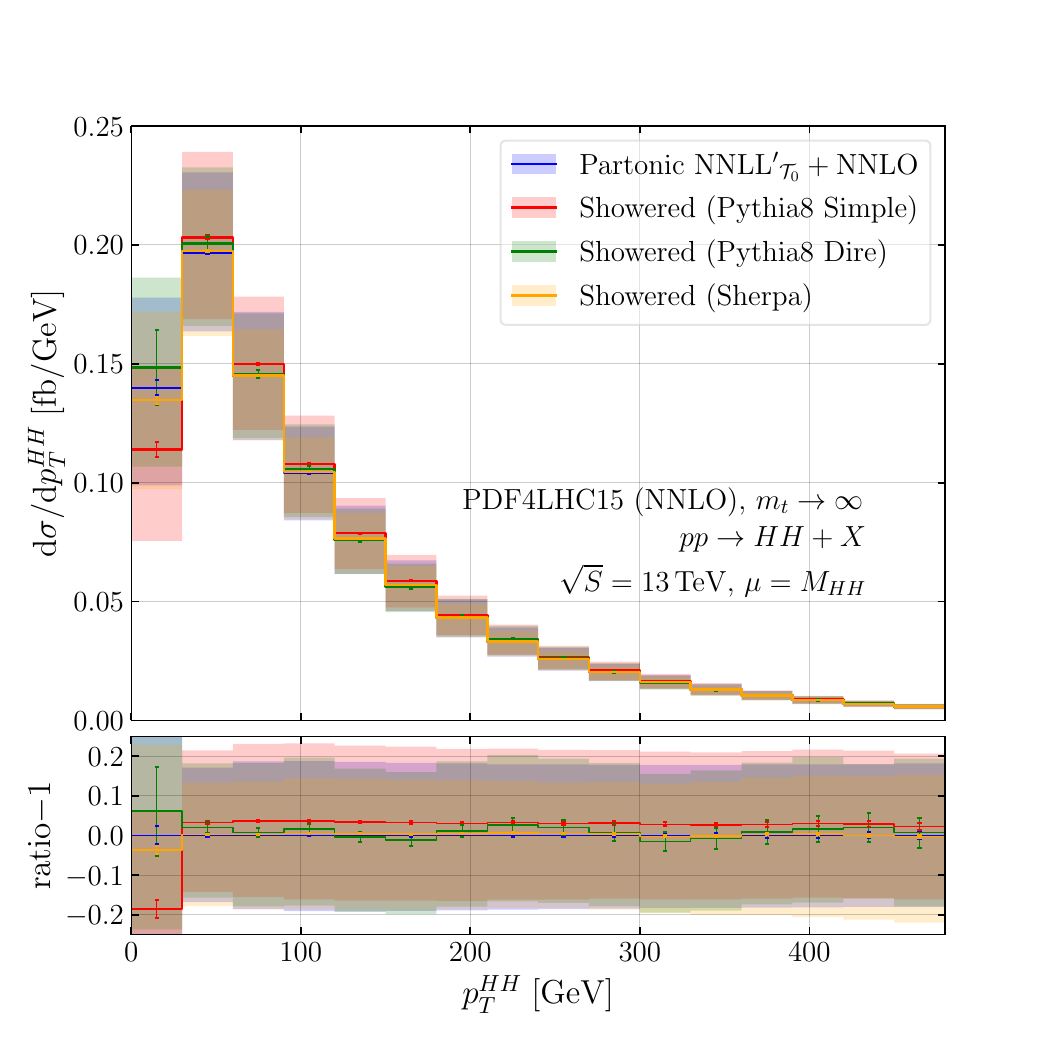}
  \end{center}
  \caption{  \label{fig:dihiggs} Comparison of the $M_{HH}$ (left) and of the $p_T^{HH}$ (right) between the partonic and showered
in \geneva, \geneva + \pythiaEight, \geneva + \dire and \geneva + \sherpa.}
\end{figure}

Considering now the double Higgs boson production, in Fig.~\ref{fig:dihiggs}, we show a comparison 
between the partonic \geneva result and the three showered results for the invariant mass (left panel) and transverse momentum of the Higgs pair (right panel).
We observe good agreement between the partonic and showered level both for inclusive and exclusive distributions.
Note also that in principle the parton shower is not required to preserve the transverse momentum, but all the shower predictions largely agree among themselfs and with the partonic result, except for the first bin where they agree with uncertainties.

Regarding these two processes presented, it would be useful to include the top-quark mass corrections to both single and double Higgs boson production.
Additionally, considering processes which involve heavy coloured partons in the final state, we are looking to the implementation of a NNLO+PS framework for the V+jets process in \geneva, with the extension of the one-jettiness resummation up to N$^3$LL.

%%%% Bibliography
\bibliographystyle{JHEP}
\bibliography{bibliography}
\end{document}